\documentclass[english,final]{elsart3p}
\usepackage[T1]{fontenc}
\usepackage[latin1]{inputenc}
\usepackage{graphicx}
\usepackage{babel}

\begin{document}
\begin{frontmatter}

\title{Application of Macro Response Monte Carlo method for electron spectrum
simulation}

\author{L. A. Perles\corauthref{quemfez}},
\ead{perles@gmail.com}
\corauth[quemfez]{Corresponding author.}
\author{A. de Almeida}
\address{Departamento de Física e Matemática\\ Faculdade de Filosofia, Ciências e Letras de Ribeirão Preto\\ São Paulo University - USP - Brazil}

\begin{abstract}
During the past years several variance reduction techniques for Monte
Carlo electron transport have been developed in order to reduce the
electron computation time transport for absorbed dose distribution.
We have implemented the Macro Response Monte Carlo (MRMC) method to
evaluate the electron spectrum which can be used used as a phase space
input for others simulation programs. Such technique uses probability
distributions for electron histories previously simulated in spheres
(called kugels). These probabilities are used to sample the primary
electron final state, as well as the creation secondary electrons
and photons. We have compared the MRMC electron spectra simulated
in homogeneous phantom against the Geant4 spectra. The results showed
an agreement better than 6\% in the spectra peak energies and that
MRMC code is up to 12 time faster than Geant4 simulations.
\end{abstract}
\end{frontmatter}

\section{Introduction}

Currently, modeling electron beam treatment planning system is a challenging
task. Several methods have been provided to calculate the tridimentional
(3D) dose distributions, from Hogstron analytical approximation~\cite{0031-9155-26-3-008}
to Monte Carlo (MC) simulations. The simulations with MC code, using
condensed history, are accurate but usually require several hours
to achieve reasonable statistics. To reduce the simulation time several
variance reduction techniques have been proposed such as Response
History Monte Carlo (RHMC)~\cite{response_history-paper}, Macro
Monte Carlo (MMC)~\cite{1992mmc,1995PMB....40..543N}, Phase-space
evolution Monte Carlo (Evolution)~\cite{scora:177}, Macro Response
Monte Carlo (MRMC)~\cite{MRMC-Phd} and Voxel Monte Carlo (VMC)~\cite{VMC}.

The RHMC, MMC, Evolution and MRMC codes use probability distributions
generated from electron histories previously simulated in a volume
element, to transport the primary electron and create secondary ones.
The Evolution generates its phase space in cubic voxels, the RHMC
generates in hemispheres and MMC and MRMC generate in spheres (kugels).
The VMC code relies on some simplifications of the models for electron
transportation and history repetition.

The cited codes have shown an accuracy up to 5\% in the depth dose,
but only RHMC, MMC and VMC have achieved a real speed gain, when compared
with others MC codes. Basically, the use of voxels instead of kugels
is the limitation of the Evolution code because it is difficult to
overcome the symmetry problem. The MRMC limitation, mentioned by the
author, was related to the small radii of the kugels used.

We have implemented the MRMC method with larger radii as well more
radii choice for electron transportation. Such implementation can
be used for generating electron spectra, since it does not have dose
deposition algorithm implemented.

\section{Methods}

\subsection{MRMC implementation}

The MRMC simulation is divided into two parts called local and global
calculations. In the first one we have simulated the kugels database
for water, soft tissue and compact bone. In the second part is the
electron interaction simulation in the materials and phantom geometries
selected.

\subsubsection{Local calculation}

\begin{figure}[!h]
\begin{centering}\includegraphics[width=0.85\columnwidth,keepaspectratio]{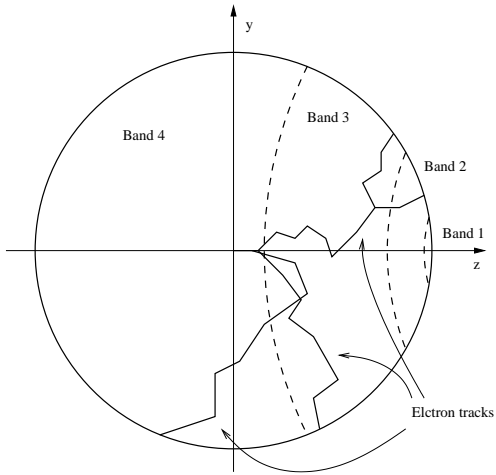}\par\end{centering}

\caption{Illustration of kugel bands with some electron tracks and its output
band.\label{fig:Ilustration-kugel-bands}}
\end{figure}

The local calculation consisted in the simulation of electron histories
starting from kugel's center with initial direction aligned to the
Z axis, figure~\ref{fig:Ilustration-kugel-bands}. The particles'
state leaving the kugel is tailed in probability distribution histograms.

The kugel is divided into 4 bands along the Z axis, where the projections
over such axis are: 1.0 to 0.98 (band~1), 0.98 to 0.8 (band~2),
0.8 to 0.4 (band~3) and 0.4 to -1.0 (band~4). Each of these bands
has a complete set of probabilities distribution histograms comprising
one for the output positions, two for directions and one for energies.
These quantities have been divided into three groups: primary electrons,
secondary electrons and photons. Also, a histogram had been generated
to store the average number of escaping secondary electrons from the
kugel geometry and another one for photons escaping.

Each material has a complete set of several kugel radii and energies.
The radii chosen were: 0.025, 0.050, 0.100, 0.250, 0.500 and 1.000~cm,
where only the first three were tested in the original work. We have
chose 19 energies stepping from 31~MeV down to 178~keV with an energy
spacing around 25\%, as suggested by Svatos~\cite{MRMC-Phd}.

We have used the Geant4 version 8.0~\cite{Geant4-NIMa,geant4-IEEE2006}
to simulate the kugels database. Such code uses a condensed history
for electron transportation, while in the original implementation
of the MRMC the author have used a Single Scattering MC code. Each
kugel has been simulated with $5\times10^{5}$ histories, using 990~eV
as electron cutoff and 10~$\mu\textrm{m}$ as rangecut for water
and soft tissue and 20~$\mu\textrm{m}$ for compact bone. Due to
long elapsed time, these simulations took place in a small cluster.

The database was stored in ROOT~\cite{ROOT-UsersGuide} file format,
which properly store kugel data for global calculations.

\subsubsection{Global calculation}

\begin{figure}[!h]
\begin{centering}\includegraphics[width=0.8\columnwidth,height=0.4\textheight]{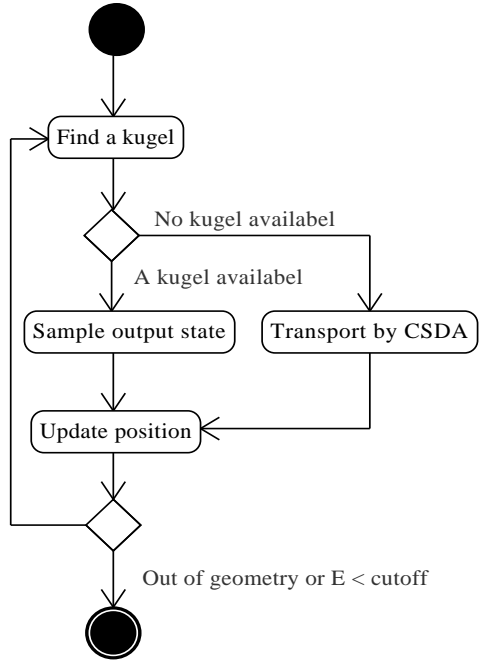}\par\end{centering}

\caption{UML Activity diagram showing the electron logic transportation.\label{fig:Activity-diagram-in}}
\end{figure}

The global calculation is the electron transport through the phantom
geometry using kugels to sample the primary electron final state,
as well the secondary electrons and photons production. An activity
diagram, in Unified Modeling Language~(UML), is presented in the
figure~\ref{fig:Activity-diagram-in}. First, the system tries to
find an energy through a linear-logarithmic statistical interpolation,
expression~\ref{eq:escala-linlog},where $E$ is the energy of electron
being transported, $E_{+}$ and $E_{-}$ are the nearest energies
available in the database and $\chi$ is an uniform random number
in $\left[0,1\right[$. The $E_{+}$ is chosen when the expression~\ref{eq:escala-linlog}
is true, otherwise $E_{-}$ will be chosen.
\begin{equation}
\chi>\frac{logE_{+}-logE}{logE_{+}-logE_{-}}\label{eq:escala-linlog}\end{equation}

After an energy is chosen, the system searches for a radius that does
not cross any boundary. If there is a kugel, the MRMC code system
uses it to sample the particles final state, including the secondaries
and photons production. If there is no kugel, the system transports
the current electron using an straight line approximation and corrects
its energy using a Continuous Slowing Down Approximation (CSDA). The
secondaries are transported as the primary ones and the photons are
transported until they leave the geometry, since interaction models
are no implemented.

The MRMC was written in object oriented style using C++ language,
with the ROOT library to sample the histograms and for the particle
transportation in the geometry.

\subsection{Benchmarks}

The tests consisted in an electron pencil beam incident perpendicularly
in a homogeneous cylindrical phantom. The electron beam starts at
0.025~cm, inside the phantom, in order to avoid the straight line
approximation in the beginning of event. The simulation parameters
for the MRMC were set to 5~keV for both, the cutoff and secondaries
threshold. The Geant4 parameters were set for 1~keV of cutoff and
0.1~mm of rangecut.

We compared the forward and lateral electron spectra scatter against
the Geant4 simulations. The energies chosen for these tests were:
10.0, 8.0, 7.5, 5.0, 3.0 and 1.0~MeV.

\section{Results}

\subsection{Electron spectra}

\begin{figure}[!h]
\begin{centering}\includegraphics[width=1\columnwidth,keepaspectratio]{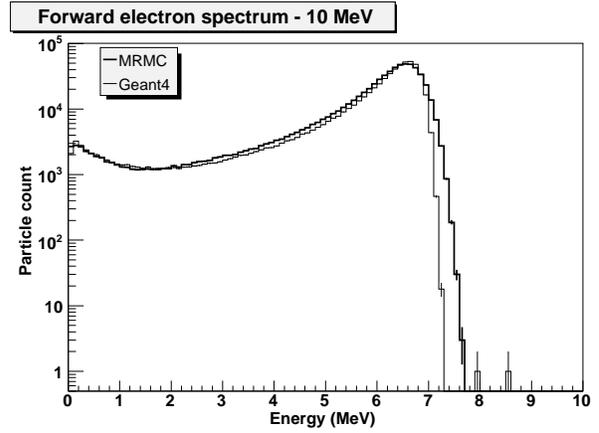}\par\end{centering}

\caption{Forward electron spectrum for incident energy of 10~MeV in water
phantom.\label{fig:Forward-electron-spectrum-h2o-10MeV}}
\end{figure}

\begin{figure}[!h]
\begin{centering}\includegraphics[width=1\columnwidth,keepaspectratio]{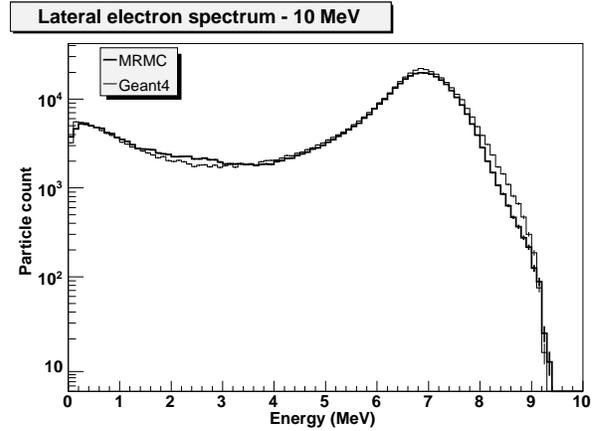}\par\end{centering}

\caption{Lateral electron spectrum for incident energy of 10~MeV in water
phantom.\label{fig:Lateral-electron-spectrum-h2o-10MeV}}
\end{figure}

\begin{figure}[!h]
\begin{centering}\includegraphics[width=1\columnwidth,keepaspectratio]{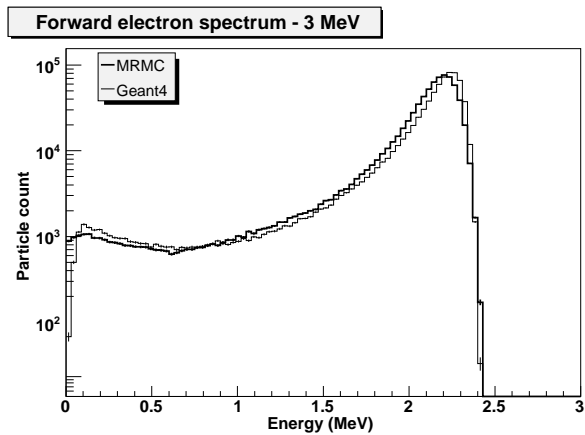}\par\end{centering}

\caption{Forward electron spectrum for incident energy of 3~MeV in soft tissue
phantom.\label{fig:Forward-electron-spectrum-tissue-3MeV}}
\end{figure}

\begin{figure}[!h]
\begin{centering}\includegraphics[width=1\columnwidth,keepaspectratio]{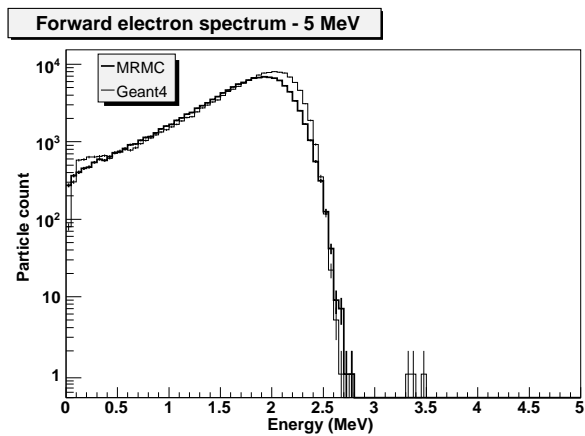}\par\end{centering}

\caption{Forward electron spectrum for incident energy of 5~MeV in compact
bone phantom.\label{fig:Forward-electron-spectrum-bone-5MeV}}
\end{figure}

\begin{figure}[!h]
\begin{centering}\includegraphics[width=1\columnwidth,keepaspectratio]{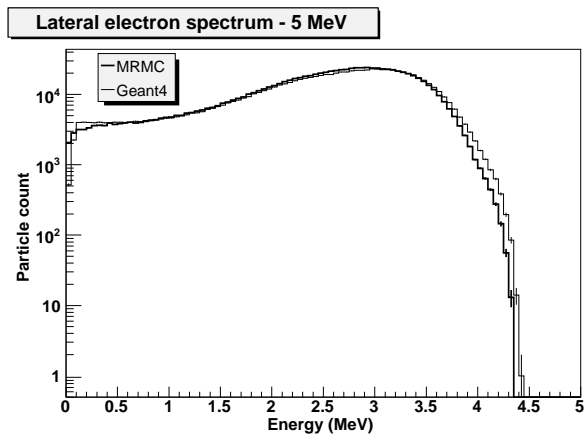}\par\end{centering}

\caption{Lateral electron spectrum for incident energy of 5~MeV in compact
bone phantom.\label{fig:Lateral-electron-spectrum-bone-5MeV}}
\end{figure}

The forward and lateral spectra simulated by MRMC for water and soft
tissue have shown an agreement better than 2\% in the peak energies
for the incident energies, above 3~MeV, and up to 5\% for incident
energies below 3~MeV, figures~\ref{fig:Forward-electron-spectrum-h2o-10MeV}
to~\ref{fig:Forward-electron-spectrum-tissue-3MeV}. For compact
bone the results showed an agreement better than 6\% for the incident
energies above 5~MeV and better than 2\% for the incident energies
below 5~MeV, figures~\ref{fig:Forward-electron-spectrum-bone-5MeV}
and~\ref{fig:Lateral-electron-spectrum-bone-5MeV}. Such deviations
may be due to uncertainties in the local simulations done with Geant4,
straight line approximation for boundary cross and uncertainties in
total stopping power tables~\cite{icru-37}.

\subsection{Simulation speed}

The MRMC simulation speed was around three times faster for lowest
energy and up to 12 times faster for 10~MeV. Simulations for low
energies are slower because the database for such energies have only
smaller kugels radii available.

\section{Conclusions}

The simulated spectra from our implementation of MRMC code have shown
good agreement and the speed was up to 12 time faster, when compared
to Geant4 for energies from 1 to 10~MeV.

The authors would like to thank to prof. Dr. Antonio Carlos Roque
da Silva Filho and Dr. Rodrigo Freire Oliveira to allow the access
to cluster for our calculations. Also, we would like to thank to CNPq
for the financial support.

\bibliographystyle{elsart-num}
\bibliography{biblio2006ago15}

\end{document}